\documentstyle[12pt]{article}
\title{Symmetric hyperbolic systems for Bianchi equations}
\author{Miguel \'A.G. Bonilla\thanks
{Also at Laboratori de F\'{\i}sica Matem\`atica,
Societat Catalana de F\'{\i}sica, IEC, Barcelona} \\
Departament de F\'{\i}sica Fonamental,
Universitat de Barcelona,\\
Av. Diagonal 647, E-08028 Barcelona, Spain\\
E-mail: mangel@ffn.ub.es}

\begin{document}

\maketitle

\begin{abstract}
We obtain a family of first-order
symmetric hyperbolic systems for the Bianchi equations.
They have only physical characteristics: the light
cone and timelike hypersurfaces. In the proof of the
hyperbolicity, new positivity properties of the Bel
tensor are used.
\end{abstract}


\section{Introduction}
First-order symmetric hyperbolic (FOSH) systems
have recently been the subject of intensive research,
not only for their theoretical interest in the Cauchy
problem, but also for their applications to numerical
relativity \cite{BOMA}. Such systems constitute a powerful
procedure that can be used to establish a well-posed initial-value
formulation for physical systems, since most of them
can be cast into this FOSH form \cite{GE} and, within this
framework, theorems of existence and uniqueness of
solutions are applicable \cite{FRI} (see also \cite{JO}).
More particularly, the Bianchi identities (investigated
for a long time mainly for its usefulness in the
study of the gravitational field propagation)
are the basis from where some FOSH systems have
been extracted (see \cite{FR} and references
therein). Many of these results
have been obtained by making use of the vacuum Bianchi
identities, however the non-vacuum ones can also be
applied to the construction of FOSH systems for the
Einstein field equations \cite{ANCH}.

In this paper a family of FOSH systems for the Bianchi
equations is constructed. They may be useful in
deriving new hyperbolic systems for Einstein's
field equations, analogously to what is done in
\cite{ANCH}, where Bianchi identities are linked
to equations for the spatial
metric and connection. We begin in section 2
by considering the Bianchi equations applied to
a double 2-form which does not satisfy {\em a priori}
the additional Riemann symmetries. From these equations
we extract a FOSH system of evolution equations
having no unphysical characteristics.
Actually, their only characteristics are the null cone
and timelike hypersurfaces. In section 3 we briefly
investigate the set of constraints that completes
the hyperbolic system found in section 2. Both together,
the evolution equations and the constraints,
are equivalent to the original Bianchi equations.
Finally, in section 4 we show that the full
Riemann symmetries (together with the Einstein
field equations) are conserved under the evolution.
In the appendix we prove new positivity properties of
the Bel tensor that are needed for the proof of
the hyperbolicity.

\section{Hyperbolizations of Bianchi equations}
We recall that the Bianchi identities for the Riemann
tensor read:
\begin{eqnarray}
\nabla_{[\nu}R_{\alpha\beta]\lambda\mu}=0 \, ,
\label{bian1riemann}
\end{eqnarray}
from where it follows (by contracting, using the
Riemann symmetries and the Einstein field equations
with $8\pi G=c=1$):
\begin{eqnarray}
\nabla^{\beta}R_{\alpha\beta\lambda\mu}=
2\nabla_{[\mu}\left(T_{\lambda]\alpha}-
\frac{1}{2}T g_{\lambda]\alpha}\right)\equiv
J_{\alpha\lambda\mu}\, ,
\label{bian2riemann}
\end{eqnarray}
$T_{\alpha\beta}$ being the energy-momentum tensor.
For the system (\ref{bian1riemann},\ref{bian2riemann}),
called the superior order field equations by Lichnerowicz,
Einstein's field equations can be
considered as initial data \cite{LI}: if they hold
at the initial spatial hypersurface $S$, they
also hold in a neighborhood of $S$.
In the last section we will see that the symmetries can
also be taken as initial data.

Consider now the set of equations
(\ref{bian1riemann},\ref{bian2riemann}), which will be
called {\em Bianchi equations} in what follows,
applied to a double 2-form
$K_{\alpha\beta\lambda\mu}=K_{[\alpha\beta][\lambda\mu]}$.
We do not suppose it to have the
additional Riemann symmetries
$K_{[\alpha\beta\lambda]\mu}=0$ at the present moment.
The Bianchi equations for
$K_{\alpha\beta\lambda\mu}$ are then:
\begin{eqnarray}
\nabla^{\beta}K_{\alpha\beta\lambda\mu}=
J_{\alpha\lambda\mu}\, ,
\label{bian1}\\
\nabla_{[\nu}K_{\alpha\beta]\lambda\mu}=0 \, . 
\label{bian2}
\end{eqnarray}
This is an overdetermined linear first-order system
of 48 equations for 36 unknowns which is not manifestly
symmetric hyperbolic. To handle more easily equations
(\ref{bian1},\ref{bian2}), it
is better to write them in the following
compact form (in a similar way to what is done in \cite{GE}
for other systems):
\begin{eqnarray}
A_{I}^{\epsilon\hspace{1mm}\gamma\delta\sigma\rho}
\nabla_{\epsilon}K_{\gamma\delta\sigma\rho}=j_{I}\, ,
\label{syst}
\end{eqnarray}
where the index $I$ lives in the space of equations,
$I=(\alpha[\lambda\mu],[\nu\alpha\beta][\lambda\mu])$, and
\begin{eqnarray*}
A_{I}^{\epsilon\hspace{1mm}\gamma\delta\sigma\rho}=
\left(\delta^{\epsilon[\delta}\delta^{\gamma]}_{\alpha}
\delta^{[\sigma}_{[\lambda}\delta^{\rho]}_{\mu]},
\delta^{\epsilon}_{[\nu}\delta^{[\gamma}_{\alpha}
\delta^{\delta]}_{\beta]}\delta^{[\sigma}_{[\lambda}
\delta^{\rho]}_{\mu]}\right)\, , \, \, \,
j_{I}=\left(J_{\alpha\lambda\mu},0\right)\, .
\end{eqnarray*}
A tensor
$H^{I\hspace{1mm}\gamma'\delta'\sigma'\rho'}$
is said to be a {\em hyperbolization} of
(\ref{syst}) if the matrix
$Q^{\epsilon\hspace{1mm}\{\gamma'\delta'\sigma'\rho'\}
\{\gamma\delta\sigma\rho\}}\equiv
H^{I\hspace{1mm}\gamma'\delta'\sigma'\rho'}
A_{I}^{\epsilon\hspace{1mm}\gamma\delta\sigma\rho}$ 
acting over the space of double 2-forms
satisfies the following properties:
{\em i)} it is symmetric, and {\em ii)} there exists a 1-form
$n_{\epsilon}$ such that
$n_{\epsilon}{\bf Q}^{\epsilon}$ is positive-definite.
By means of such a hyperbolization, we can transform
(\ref{syst}) into:
\begin{eqnarray*}
H^{I\hspace{1mm}\gamma\delta\sigma\rho}
A_{I}^{\epsilon\hspace{1mm}\gamma'\delta'\sigma'\rho'}
\nabla_{\epsilon}K_{\gamma'\delta'\sigma'\rho'}=
H^{I\hspace{1mm}\gamma\delta\sigma\rho}j_{I} \, ,
\end{eqnarray*}
which is now a FOSH system and, therefore, it admits the
theorems of existence and uniqueness of solutions.

The set of 1-forms $n_{\epsilon}$ satisfying property
{\em ii)} has a clear physical meaning (see \cite{GE}
for instance) related to the propagation of the physical
fields involved and to the causal character of the system.
In fact, a FOSH system is said to be {\em causal} if every
$n_{\epsilon}$ satisfying {\em ii)} is a future-directed
1-form.

For the Bianchi equations (\ref{syst}) the following
family of tensors are hyperbolizations:
\begin{eqnarray*}
H^{I\hspace{1mm}\gamma'\delta'\sigma'\rho'}=
\left(\delta^{\alpha[\gamma'}
t^{\delta']}h^{[\lambda\mu][\sigma'\rho']},
-\frac{3}{2}t^{[\nu}\delta^{\alpha \,\mid [\gamma'}
\delta^{\delta']\mid\, \beta]}
h^{[\lambda\mu][\sigma'\rho']}\right) \, ,\\
h^{\lambda\mu\sigma'\rho'}=g^{\lambda\sigma'}
\left[u^{\left(\mu\right.}v^{\left.\rho'\right)}
-\frac{1}{4}
\left(u_{\nu}v^{\nu}\right)g^{\mu\rho'}\right]\, ,
\end{eqnarray*}
where $\vec{t}$, $\vec{u}$ and $\vec{v}$ are arbitrary
timelike future-directed vectors. 
To check property {\em i)} we compute, for any two
double 2-forms $S_{\gamma\delta\sigma\rho}$ and
$T_{\gamma\delta\sigma\rho}$, the quantity
\begin{eqnarray*}
Q^{\epsilon\hspace{1mm}\{\gamma'\delta'\sigma'\rho'\}
\{\gamma\delta\sigma\rho\}}
S_{\gamma'\delta'\sigma'\rho'}
T_{\gamma\delta\sigma\rho}= \\
=\left[
\frac{1}{2}\left(S^{\alpha\hspace{1mm}\beta}
_{\hspace{2mm}\nu\hspace{2mm}\sigma}
T_{\alpha\mu\beta\rho}+
T^{\alpha\hspace{1mm}\beta}
_{\hspace{2mm}\nu\hspace{2mm}\sigma}
S_{\alpha\mu\beta\rho}+
S^{\alpha\hspace{1mm}\beta}_{\hspace{2mm}
\nu\hspace{2mm}\rho}
T_{\alpha\mu\beta\sigma}+
T^{\alpha\hspace{1mm}\beta}_{\hspace{2mm}
\nu\hspace{2mm}\rho}
S_{\alpha\mu\beta\sigma}\right)- \right. \\
\left.
-\frac{1}{4}g_{\nu\mu}
\left(S^{\alpha\beta\gamma}_{\hspace{5mm}\sigma}
T_{\alpha\beta\gamma\rho}+
T^{\alpha\beta\gamma}_{\hspace{5mm}\sigma}
S_{\alpha\beta\gamma\rho}\right)-
\frac{1}{4}g_{\sigma\rho}
\left(S_{\nu}^{\hspace{2mm}\alpha\beta\gamma}
T_{\mu\alpha\beta\gamma}+
T_{\nu}^{\hspace{2mm}\alpha\beta\gamma}
S_{\mu\alpha\beta\gamma}\right)+ \right.\\
\left.
+\frac{1}{8}g_{\nu\mu}g_{\sigma\rho}
S_{\alpha\beta\gamma\delta}
T^{\alpha\beta\gamma\delta}\right]
g^{\nu\epsilon}t^{\mu}u^{\sigma}v^{\rho}=
Q^{\epsilon\hspace{1mm}\{\gamma'\delta'\sigma'\rho'\}
\{\gamma\delta\sigma\rho\}}
T_{\gamma'\delta'\sigma'\rho'}
S_{\gamma\delta\sigma\rho} \, ,
\end{eqnarray*}
from where the symmetry is clear.
Property {\em ii)} is demonstrated in the appendix
by proving that $n_{\epsilon}
{\bf Q}^{\epsilon}$ is a positive-definite
quadratic form for any $n_{\epsilon}$ timelike
future-directed 1-form. This also shows that
the system is causal.

Let us now study the characteristics of the
hyperbolized Bianchi equations found above,
which finally read:
\begin{eqnarray}
h_{\alpha\beta}^{\hspace{4mm}[\sigma\rho]}
\left(t^{[\delta}\nabla_{\nu}K^{\gamma]\nu\alpha\beta}-
\frac{3}{2}t_{\epsilon}\nabla^{[\epsilon}
K^{\gamma\delta]\alpha\beta}\right)=
h_{\alpha\beta}^{\hspace{4mm}[\sigma\rho]}
\left(t^{[\delta}J^{\gamma]\alpha\beta}\right) \, .
\label{foshs}
\end{eqnarray}
Recall that a 1-form $\zeta_{\alpha}$ is, by definition,
{\em characteristic} for the system (\ref{foshs}) if
$\det(\zeta_{\epsilon}{\bf Q}^{\epsilon})=0$. To
study the kernel of the characteristic matrix,
first notice that $h_{[\alpha\beta][\lambda\mu]}$
(considered as a matrix acting over 2-forms)
has non-zero determinant, that is:
\begin{eqnarray}
h_{[\alpha\beta][\lambda\mu]}
A^{\lambda\mu}=0 \, \Longrightarrow \,
A^{[\lambda\mu]}=0 \, .
\label{nonzerodet}
\end{eqnarray}
This can be easily proven by contracting
(\ref{nonzerodet}) successively with $\vec{u}$ and
$\vec{v}$. Therefore,
\begin{eqnarray*}
Q^{\epsilon\hspace{1mm}\{\gamma'\delta'\sigma'\rho'\}
\{\gamma\delta\sigma\rho\}}
\zeta_{\epsilon}K_{\gamma\delta\sigma\rho}=0 \, \,
\Longleftrightarrow \, \,
t^{[\delta}\zeta_{\nu}K^{\gamma]\nu\alpha\beta}-
\frac{3}{2}t_{\epsilon}\zeta^{[\epsilon}
K^{\gamma\delta]\alpha\beta}=0 \, .
\end{eqnarray*}
The non-trivial solutions of this system
are found to be of two types:
{\em i) $\zeta_{\alpha}\zeta^{\alpha}=0$} and
{\em ii) $\zeta_{\alpha}t^{\alpha}=0$}.
In the first case, the solutions
are the double 2-forms $K_{\alpha\beta\lambda\mu}$
satisfying $\zeta^{\alpha}K_{\alpha\beta\lambda\mu}=0$
and $\zeta_{[\nu}K_{\alpha\beta]\lambda\mu}=0$.
In the second one the solutions are those
$K_{\alpha\beta\lambda\mu}$ having in its two first
indices the structure $A\left(t\wedge\zeta\right)+
B *\left(t\wedge\zeta\right)$
(``$\ast$'' is the usual dual operator).
Therefore, as remarked above, (\ref{foshs}) has
only physical characteristics: {\em i)} the light
cone and {\em ii)} timelike hypersurfaces.

\section{Equivalence with the original system}
From the Bianchi equations we have constructed
the 36 propagation equations (\ref{foshs})
(in the sense that they contain derivatives in
any timelike direction $n_{\alpha}$). In this
section we will show that these equations, when
completed with the constraints, are equivalent to
the original system (\ref{bian1},\ref{bian2}).
Taking $n_{\nu}$ as the normal
to an initial-value spatial hypersurface $S$ and
$\eta^{\epsilon\alpha\beta\nu}$ the spacetime
volume 4-form, the constraints are:
\begin{eqnarray}
n^{\alpha}\left(\nabla^{\beta}K_{\alpha\beta\lambda\mu}-
J_{\alpha\lambda\mu}
\right)=0 \, ,
\label{const1}\\
n_{\nu}\eta^{\epsilon\alpha\beta\nu}
\nabla_{\epsilon}K_{\alpha\beta\lambda\mu}=0 \, ,
\label{const2}
\end{eqnarray}
where all the derivatives appearing in these expressions
are tangent to $S$. This set of constraints is complete
(there are 12 equations) and also integrable provided that
$\nabla^{\alpha}J_{\alpha\lambda\mu}=0$, which
holds because of the Einstein field equations.

Taking into account (\ref{nonzerodet}),
equation (\ref{foshs}) immediately implies
\begin{eqnarray}
t^{[\delta}\nabla_{\nu}K^{\gamma]\nu\alpha\beta}-
\frac{3}{2}t_{\epsilon}
\nabla^{[\epsilon}
K^{\gamma\delta]\alpha\beta}-
t^{[\delta}J^{\gamma]\alpha\beta}=0 \, .
\label{eqconst}
\end{eqnarray}
If we multiply the previous expression by $t_{\delta}$,
we obtain $P^{\sigma}_{\alpha}
\left(\nabla^{\beta}K_{\sigma\beta\lambda\mu}-
J_{\sigma\lambda\mu}\right)=0$, where
$P^{\sigma}_{\alpha}$ is the orthogonal projector
to $t_{\alpha}$. Taking into account (\ref{const1})
and the fact that $t_{\alpha}n^{\alpha}\neq0$, we
obtain equation (\ref{bian1}).
Equation (\ref{bian2}) is achieved in a similar
way: we first multiply (\ref{eqconst}) by
$\eta_{\gamma\delta\sigma\rho}t^{\rho}$
obtaining
$P^{\sigma}_{\nu}\eta^{\epsilon\alpha\beta\nu}
\nabla_{\epsilon}K_{\alpha\beta\lambda\mu}=0$,
and then we follow the same reasoning as before,
but now making use of constraint (\ref{const2}).
This shows that the set of equations
(\ref{foshs},\ref{const1},\ref{const2})
is equivalent to (\ref{bian1},\ref{bian2}).

\section{Riemann symmetries are conserved}
Until now, we have only dealt with double 2-forms,
but, actually, we are interested in the
Riemann tensor. Our purpose now is to show that
Bianchi equations preserve the full Riemann
symmetries.
The complete set of Riemann symmetries is
$K_{\alpha\beta\lambda\mu}=K_{[\alpha\beta][\lambda\mu]}$
and $K_{[\alpha\beta\lambda]\mu}=0$.
In order to construct an appropriate system of
differential equations, let us set the following
definitions:
\begin{eqnarray*}
\Xi_{\alpha\beta\lambda\mu}\equiv
K_{[\alpha\beta\lambda]\mu}\, , \; \; 
\Sigma_{\alpha\beta}\equiv
K_{\alpha\beta}-\left(T_{\alpha\beta}-
\frac{1}{2}T g_{\alpha\beta}\right) \, ,
\end{eqnarray*}
where $K_{\alpha\beta}\equiv
K^{\sigma}_{\hspace{1mm}\alpha\sigma\beta}$.
Then we extract, from equations
(\ref{bian1},\ref{bian2})
and the energy-momentum conservation equation
$\nabla^{\beta}T_{\alpha\beta}=0$,
the following system for the previous defined variables
$\Xi_{\alpha\beta\lambda\mu}$ and $\Sigma_{\alpha\beta}$:
\begin{eqnarray}
\nabla^{\alpha}\Xi_{\alpha\beta\lambda\mu}
+\frac{2}{3}\nabla_{[\beta}\Sigma_{\lambda]\mu}=0\, ,
\nonumber\\
\nabla^{\alpha}\Sigma_{\alpha\beta}=0\, ,
\nonumber\\
\nabla^{\beta}\Sigma_{\alpha\beta}
-\frac{1}{2}\nabla_{\alpha}\Sigma=0 \, .
\label{symm}
\end{eqnarray}
These are 32 equations for 32 unknowns forming
a linear first-order homogeneous system whose only
characteristic is the null cone. Now, we suppose that
initially the double 2-form
$K_{\alpha\beta\lambda\mu}$
satisfies the full Riemann symmetries
($\Xi_{\alpha\beta\lambda\mu}=0$) and that,
in addition, $K_{\alpha\beta}$ satisfies
Einstein's equations in the sense $\Sigma_{\alpha\beta}=0$.
These initial conditions have no other solution
but $\Xi_{\alpha\beta\lambda\mu}=\Sigma_{\alpha\beta}=0$
(see \cite{JO}, for instance) and,
therefore, the Riemann symmetries, likewise the
Einstein field equations, are preserved by the
Bianchi equations.

\section*{Acknowledgments}
We wish to thank Carlos F. Sopuerta and Sergi R.
Hildebrandt for a careful reading of this manuscript.
We also acknowledge the {\it Comissionat per a
Universitats i Recerca de la Generalitat de Catalunya}
for financial support.

\appendix
\section{Positivity property of Bel tensor}
To prove that $n_{\epsilon}
{\bf Q}^{\epsilon}$ is a positive-definite
quadratic form we will make use of the standard spinor
calculus \cite{PERI}. First, taking into account
the symmetries of $K_{\alpha\beta\lambda\mu}$, we
decompose it as follows:
\begin{eqnarray*}
K_{\nu\mu\sigma\rho}=
\chi_{NMSR}\epsilon_{N'M'}\epsilon_{S'R'}
+\Psi_{NMS'R'}\epsilon_{N'M'}\epsilon_{SR} \\
+\overline{\chi}_{N'M'S'R'}\epsilon_{NM}\epsilon_{SR}+
\overline{\Psi}_{N'M'SR}\epsilon_{NM}\epsilon_{S'R'} \, ,
\end{eqnarray*}
where
\begin{eqnarray*}
\chi_{NMSR}=\chi_{(NM)(SR)}\, ,
\Psi_{NMS'R'}=\Psi_{(NM)(S'R')} \, .
\end{eqnarray*}
Now we proceed to calculate 
\begin{eqnarray}
n_{\epsilon}Q^{\epsilon\hspace{1mm}
\{\gamma'\delta'\sigma'\rho'\}
\{\gamma\delta\sigma\rho\}}
K_{\gamma\delta\sigma\rho}K_{\gamma'\delta'\sigma'\rho'}=
\left(K^{\alpha\hspace{1mm}\beta}
_{\hspace{2mm}\nu\hspace{2mm}\sigma}
K_{\alpha\mu\beta\rho}+
K^{\alpha\hspace{1mm}\beta}_{\hspace{2mm}\nu\hspace{2mm}\rho}
K_{\alpha\mu\beta\sigma}- \right. 
\nonumber
\\
\left.
-\frac{1}{2}g_{\nu\mu}
K^{\alpha\beta\gamma}_{\hspace{5mm}\sigma}
K_{\alpha\beta\gamma\rho}-
\frac{1}{2}g_{\sigma\rho}
K_{\nu}^{\hspace{1mm}\alpha\beta\gamma}
K_{\mu\alpha\beta\gamma}+
\frac{1}{8}g_{\nu\mu}g_{\sigma\rho}
K_{\alpha\beta\gamma\delta}
K^{\alpha\beta\gamma\delta}\right)
n^{\nu}t^{\mu}u^{\sigma}v^{\rho} \, ,
\label{bel}
\end{eqnarray}
which is the Bel tensor \cite{BEL} for the double 2-form
$K_{\alpha\beta\gamma\delta}$ contracted
with four timelike future-directed vectors,
$\vec{n}$, $\vec{t}$, $\vec{u}$ and $\vec{v}$.
In fact, this is a little more general since we have
not required symmetry between indices $\alpha\beta$
and $\lambda\mu$. In order
to express (\ref{bel}) in spinor form it is more
convenient to rewrite it in the following form:
\begin{eqnarray*}
\frac{1}{2}\left(
K^{\alpha\hspace{1mm}\beta}_{\hspace{2mm}
\nu\hspace{2mm}\sigma}
K_{\alpha\mu\beta\rho}+   
{K*}^{\alpha\hspace{1mm}\beta}_{\hspace{2mm}
\nu\hspace{2mm}\sigma}
{K*}_{\alpha\mu\beta\rho}+ \right. \\
\left.   
+{*K}^{\alpha\hspace{1mm}\beta}_{\hspace{2mm}
\nu\hspace{2mm}\sigma}
{*K}_{\alpha\mu\beta\rho}+
{*K*}^{\alpha\hspace{1mm}\beta}_{\hspace{2mm}
\nu\hspace{2mm}\sigma}
{*K*}_{\alpha\mu\beta\rho}\right)
n^{\nu}t^{\mu}u^{\sigma}v^{\rho} \, ,
\end{eqnarray*}
and now, the dual terms appearing in this expression have
simple spinor expressions (see \cite{PERI}).
Performing the calculations we finally obtain
\begin{eqnarray}
n_{\epsilon}Q^{\epsilon\hspace{1mm}
\{\gamma\delta\sigma\rho\}
\{\gamma'\delta'\sigma'\rho'\}}
K_{\gamma\delta\sigma\rho}K_{\gamma'\delta'\sigma'\rho'}= 
\nonumber\\
=4\left(\chi_{NMSR}\overline{\chi}_{N'M'S'R'}+
\Psi_{NMS'R'}\overline{\Psi}_{N'M'SR}\right)
n^{\nu}t^{\mu}u^{\sigma}v^{\rho} \, ,
\label{posit}
\end{eqnarray}
which is always positive, and zero if and only if
$\chi_{NMSR}=\Psi_{NMS'R'}=0$. This is proven
by using usual methods (see, for instance,
\cite{PERI,BE}): let us consider a spinor basis $\left(\iota_{A},
o_{A}\right)$ and its two associated null
future-directed vectors, 
$l_{\alpha}=\iota_{A}\overline{\iota}_{A'}$ and
$k_{\alpha}=o_{A}\overline{o}_{A'}$.
Given a timelike future-pointing vector
$\vec{z}$ and a future-directed null vector $\vec{x}$,
there exists another future-directed null vector
$\vec{y}$ such that: 
\begin{eqnarray*}
\vec{z}=a\vec{x}+\vec{y}=a
x_{A}\overline{x}_{A'}+
y_{A}\overline{y}_{A'} \, , \, \,
\mbox{$a>0$.}
\end{eqnarray*}
Applying this decomposition to
$\vec{n}$, $\vec{t}$, $\vec{u}$ and $\vec{v}$,
(\ref{posit}) becomes a sum of positive terms,
so the positivity is shown.
Suppose now that the expression (\ref{posit}) is zero.
We can decompose the four previous vectors in many
differents ways by choosing their corresponding
vector $\vec{x}$ to be $\vec{l}$ or $\vec{k}$.
Then, for every choice we do, among
the positive terms of the sum we will get a component
of $\chi_{NMSR}$ and $\Psi_{NMS'R'}$ in the
spinor basis mentioned above. Therefore, making all
the possible choices, we will obtain all the
components zero, which means that
$K_{\alpha\beta\lambda\mu}=0$.

\end{document}